\newif\ifAbstract
\Abstractfalse
\newif\ifConf
\Conftrue

\ifConf
\documentclass{sig-alternate}
\else
\documentclass[11pt]{article}
\usepackage{fullpage}
\fi

\usepackage{graphicx}
\newtheorem{theorem}{Theorem}
\newtheorem{lemma}[theorem]{Lemma}

\newcommand{\sq}{\hbox{\rlap{$\sqcap$}$\sqcup$}}
\ifConf
\else
\newcommand{\qed}{\hspace*{\fill}\sq}
\newenvironment{proof}{\noindent {\bf Proof.}\ }{\qed\par\vskip 4mm\par}
\fi

\newcommand{\ignore}[1]{ }

\def\R{\ensuremath{\mathcal{R}}}
\def\P{\ensuremath{\mathcal{P}}}
\def\C{\ensuremath{\mathcal{C}}}

\begin{document}

\ifConf
\conferenceinfo{PODC'05,}{July 17--20, Las Vegas, Nevada, USA.}
\CopyrightYear{2005}
\crdata{1-59593-994-2/05/0007}

\ifAbstract
\setcounter{page}{0}
\thispagestyle{empty}
\fi

\title{
{\huge {Skip-Webs: Efficient Distributed Data Structures}} \\[5pt]
{\huge {for Multi-Dimensional Data Sets}}
}
\date{}
\ifAbstract
\author{
 Lars Arge\thanks{Dept.~of Computer Science, University of Aarhus, 
	  IT-Parken, Aabogade 34, DK-8200 Aarhus N, Denmark.
	  \texttt{large(at)daimi.au.dk}.}
\and
David Eppstein\thanks{
	Dept.~of Computer Science, University of California,
	Irvine, CA 92697-3425, USA.  \texttt{eppstein(at)ics.uci.edu.}}
\and
Michael T. Goodrich\thanks{
	Dept.~of Computer Science, University of California,
	Irvine, CA 92697-3425, USA.  \texttt{goodrich(at)acm.org.}}
}
\else
\author{\rm
 \begin{tabular}{ccc}
 {\large\bfseries Lars Arge} & {\large\bfseries David Eppstein} & 
 {\large\bfseries Michael T. Goodrich} \\[4pt]
 Dept.~of Computer Science &
 Dept.~of Computer Science &
 Dept.~of Computer Science \\
 University of Aarhus &
	University of California &
	University of California \\
	  IT-Parken, Aabogade 34 &
	  Computer Science Bldg., 444 &
	  Computer Science Bldg., 444 \\
	  DK-8200 Aarhus N, Denmark &
	Irvine, CA 92697-3425, USA  &
	Irvine, CA 92697-3425, USA  \\[2pt]
	  \texttt{large(at)daimi.au.dk} &
	\texttt{eppstein(at)ics.uci.edu} &
	\texttt{goodrich(at)acm.org}
  \end{tabular}
}
\fi
\maketitle
\thispagestyle{empty}

\begin{abstract}
We present a framework for designing 
efficient distributed data structures for multi-dimensional data.
Our structures, which we call \emph{skip-webs}, extend and improve previous
randomized distributed data structures, including skipnets and skip graphs.
Our framework applies to a general class of data querying scenarios, which
include linear (one-dimensional) data, such as sorted sets,
as well as multi-dimensional data, such as $d$-dimensional octrees
and
digital tries of character strings defined over a fixed alphabet.
We show how to perform a query over such a set of $n$ items 
spread among $n$ hosts using $O(\log n/\log\log n)$ messages for
one-dimensional data, or
$O(\log n)$ messages for fixed-dimensional data, while using
only $O(\log n)$
space per host.  We also show how to make such structures dynamic so as to
allow for insertions and deletions in $O(\log n)$ messages
for quadtrees, octrees, and digital tries, and
$O(\log n/\log\log n)$ messages for one-dimensional data.
Finally, we 
show how to apply a blocking strategy to skip-webs to further improve message
complexity for one-dimensional data 
when hosts can store more data.


\ifAbstract
\medskip
\noindent\textbf{Number of pages:} 10

\medskip
\noindent\textbf{Category:} Regular paper; no student authors; not to be
considered for brief announcement
\fi
\end{abstract}

\ifAbstract
\bigskip
\begin{center}
\textbf{Contact author:} \\[5pt]
Michael T. Goodrich \\
Dept. of Computer Science \\
Computer Science Bldg., 444 \\
University of California, Irvine \\
Irvine, CA 92697-3425 \\
Phone: 949-824-9366 \\
Fax: 208-693-4984 \\
Email: goodrich(at)acm.org
\end{center}
\fi

\ifAbstract
\clearpage
\fi
\section{Introduction} 
\label{sec:introduction}
Peer-to-peer networks offer a decentralized, distributed way of
storing large data sets.
They store
data at the hosts in a network and they allow searches
to be implemented by
sending messages between hosts, so as to route a query to the host
that stores the requested information.
That is, data is stored at the nodes of a network according to some
indexing strategy organized over a set of data attributes, 
with the desire that users should be able to
quickly access this data using attribute-based queries.
In this paper, we are interested in allowing for a fairly
rich set of possible data queries, including
an exact match for a key (e.g., a file name), a
prefix match for a key string, a nearest-neighbor match for a
numerical attribute,
a range query over various numerical attributes, or 
a point-location query in a geometric map of attributes.
That is, we would like the peer-to-peer network to support a rich set
of possible data types that allow for multiple kinds of queries, 
including set membership, 1-dim.~nearest neighbor
queries, range queries, string prefix queries, and
point-location queries.

The motivation for such queries include DNA databases, location-based
services, and approximate searches for file names or data titles.
For example, a prefix query for ISBN numbers in a book database could return
all titles by a certain publisher.
Likewise, a nearest-neighbor query in a two-dimensional point set could
reveal the closest open computer kiosk or empty parking space
on a college campus.
By allowing for such databases to be stored, searched, and updated in a
distributed peer-to-peer (p2p) network, application providers can maximize
information availability and accuracy at low cost to any individual
host participant (assuming honest participants, of course\footnote{%
  The security of distributed data structures is an interesting 
  research topic~\cite{agt-pada-01,gttc-adsggs-03,xor-asdss-02},
  but is beyond the scope of this paper.}).

The key challenges in setting up such p2p networks is to balance
the space and processing loads of the hosts while providing for low
message costs for routing queries and performing 
insertion and deletion updates in the data set.
We would like the data to be distributed as evenly as possible across
the network, so that the space-usage load of each host is roughly the
same.
And we desire that the data management be fully decentralized, with hosts
able to insert and delete data they own without employing a centralized data
manager or repository (which would create a single point of failure).
Likewise, we would like the indexing structure to be distributed in
such a way that the query-processing load is also spread as uniformly
as possible across the nodes of the network.
This paper is therefore directed at the study of distributed peer-to-peer
data structures for multi-dimensional data.

\begin{figure*}[tb]
\centerline{\includegraphics[scale=0.85]{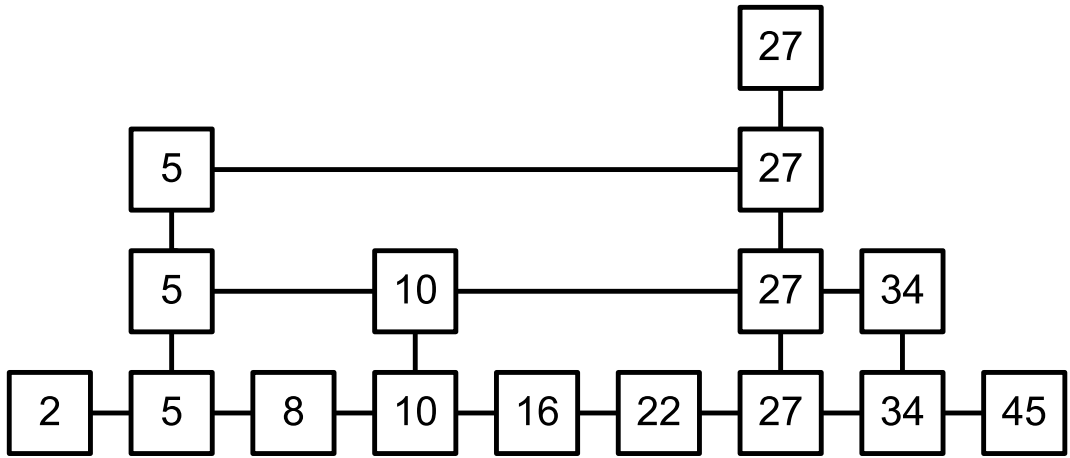}}
\caption{A skip list data structure.  Each element exists in the
bottom-level list, and each node on one level is copied to the next
higher level with probability $1/2$.  A search starts at the top and
proceeds as far as it can on a given level, then drops down to the
next level, and continues until it reaches the desired node on the
bottom level.  The expected query time is $O(\log n)$ and the
expected space is $O(n)$.}
\label{fig:skiplist}
\end{figure*}

\subsection{Cost Measures and Models for Peer-to-Peer Data Structures}

There are a number of cost measures for quantifying the quality of a
distributed data structure for an p2p network.
First and foremost, we have the following two parameters:
\begin{itemize}
\item
$H$: the number of hosts.  
Each host is a computer on the network with a unique ID. 
We assume the network allows
any host to send a message to any other host.
We also assume that hosts do not fail.
\item
$M$: the maximum memory size of a host.  
The memory size is measured by the number
of data items, data structure nodes, pointers, and host IDs that any host can
store.  
\end{itemize}

In addition to the parameters $H$ and $M$, we have the following
cost functions for a distributed data structure storing a set $S$
of $n$ items in a peer-to-peer network:
\begin{itemize}
\item
$Q(n)$: the \emph{query cost}---the number of messages
needed to process a query on $S$.
\item
$U(n)$: the \emph{update cost}---the number of messages needed to
insert a new item in the set $S$ or remove an item from the
set $S$.
\item 
$C(n)$: the \emph{congestion} per host---the
sum of the number of references to items stored at the host,
the number of references to items stored at other hosts,
and the number $n/H$ (which measures the expected number of queries likely to
begin at any host, based on the number of items in the set $S$).
\end{itemize}
We assume that each host has a reference to the place where any
search from that host should begin, i.e., a \emph{root node} for that
host (which may be stored on the host itself).

\subsection{Previous Related Work}
We are not aware of any previous related work on storing multi-dimensional
data in peer-to-peer networks.

Nevertheless,
there are existing efficient structures for storing and retrieving
one-dimensional data in peer-to-peer networks.
For example, there is a considerable amount of
prior work on variants of 
Distributed Hash Tables (DHTs), 
including Chord~\cite{gm-orc-04,smkkb-csppl-01}, 
Koorde~\cite{kk-asdod-03},
Pastry~\cite{rowstron01pastry},
Scribe~\cite{rowstron01scribe},
Symphony~\cite{mbr-sdsw-03},
and 
Tapestry~\cite{zhao01tapestry}, to name just a few.
These structures do not allow for sophisticated queries, however,
such as nearest-neighbor searching (even in one-dimension), 
string prefix searching,
or multi-dimensional point location.

For one-dimensional nearest-neighbor searching,
Aspnes and Shah~\cite{as-sg-03}
present a distributed data structure, called \emph{skip graphs}, for
searching ordered content in a peer-to-peer network,
based on the elegant, randomized skip-list data structure~\cite{p-slpab-90}.
(See Figure~\ref{fig:skiplist}.)
Harvey \textit{et al.}~\cite{hjstw-ssonp-03} independently present a
similar structure, which they call SkipNet.
These structures achieve $O(\log n)$ congestion, expected query time,
and expected update times, using $n$ hosts.
Harvey and Munro~\cite{hm-ds-03} present a deterministic version of
SkipNet, showing how updates can be performed efficiently, albeit
with higher congestion and update costs, which are $O(\log^2 n)$.
Zatlukal and Harvey~\cite{zh-ftaod-04}
show how to modify SkipNet, to construct a
structure they call family trees, to achieve $O(\log n)$ expected
time for search and update, while restricting the number of pointers 
from one host to other hosts to be $O(1)$.
Manku, Naor, and Wieder~\cite{mnw-knnpl-04} show how to improve the
expected 
query cost for searching skip graphs and SkipNet to $O(\log n/\log\log n)$, 
by having hosts store the pointers from their neighbors to their
neighbor's neighbors (i.e., neighbors-of-neighbors (NoN) tables);
see also Naor and Wieder~\cite{nw-knnbr-04}.
Unfortunately, this improvement requires that the memory size,
congestion, and expected update time grows to be $O(\log^2 n)$.
Focusing instead on fault tolerance,
Awerbuch and Scheideler~\cite{awerbuch03peertopeer} show how to combine 
a skip graph/SkipNet data structure with a DHT to achieve improved
fault tolerance\footnote{Achieving improved fault tolerance in
	peer-to-peer data structures for multi-dimensional data is a topic
	for possible future research.}, 
but at an expense of a logarithmic factor slow-down
for queries and updates.
Aspnes \textit{et al.}~\cite{akk-lblrq-04} show how to reduce the
space complexity of the skip graphs structure, by bucketing intervals
of keys on the ``bottom level'' of their structure.  Their method
improves the expected bounds for searching and updating, but only by
a constant factor whenever $H$ is $\Theta(n^{\epsilon})$, for constant
$\epsilon>0$.

\begin{table*}[tb]
\begin{center}
\footnotesize
\begin{tabular}{|c|c|c|c|c|c|}
\hline\hline
\emph{Method} & $H$ & $M$ & $C(n)$ & $Q(n)$ & $U(n)$ \\
\hline\hline
skip graphs/SkipNet~\cite{as-sg-03,hjstw-ssonp-03} &
    $n$ & $O(\log n)$ & $O(\log n)$ & $\tilde O(\log n)$ & $\tilde O(\log n)$ 
    \rule{0pt}{11pt}\\
\hline
NoN skip-graphs~\cite{mnw-knnpl-04,nw-knnbr-04} &
 $n$ & $O(\log^2 n)$ & $O(\log^2 n)$ & $\tilde O(\log n/\log\log n)$ & 
                        $\tilde O(\log^2 n)$ 
			\rule{0pt}{11pt}\\
\hline
family trees~\cite{zh-ftaod-04} &
 $n$ & $O(1)$ & $O(\log n)$ & $\tilde O(\log n)$ & $\tilde O(\log n)$ 
 \rule{0pt}{11pt}\\
\hline
deterministic SkipNet~\cite{hm-ds-03} &
 $n$ & $O(\log n)$ & $O(\log n)$ & $O(\log n)$ & $O(\log^2 n)$ 
 \rule{0pt}{11pt}\\
\hline
bucket skip graphs~\cite{akk-lblrq-04} &
 $<n$ & $O(n/H + \log H)$ & $O(n/H + \log H)$ & $\tilde O(\log H)$ 
       & $\tilde O(\log H)$ 
       \rule{0pt}{11pt}\\
\hline
\hline
\textbf{skip-webs} &
  $n$ & $O(\log n)$ & $O(\log n)$ & $\tilde O(\log n/\log\log n)$ & 
        $\tilde O(\log n/\log\log n)$ 
	\rule{0pt}{11pt}\\
\hline
\textbf{bucket skip-webs} &
  $<n$ & $O(n/H+\log H)$ & $O(n/H + \log H)$ & $\tilde O(\log_M H)$
        & $\tilde O(\log_M H)$ 
	\rule{0pt}{11pt}\\
\hline\hline
\end{tabular}
\end{center}
\caption{Comparison of 1-dimensional skip-webs with previous methods
for 1-dimensional nearest neighbor structures.
We use $\tilde O(*)$ to
denote an expected cost bound. The skip-webs and bucket skip-webs
solutions are presented in this paper.}
\label{tbl-compare}
\end{table*}

\pagebreak
\subsection{Our Results}
In this paper, we present a framework for designing 
randomized distributed data structure
that improves previous skip-graph/SkipNet approaches and extends
their area of applicability to multi-dimensional data sets.
The queries allowed
include one-dimensional nearest neighbor queries, string searching over
fixed alphabets, and multi-dimensional searching and point location.
Our structure, which we call \emph{skip-webs}, matches
the $O(\log n/\log \log n)$ expected query time 
of NoN skip-graphs~\cite{mnw-knnpl-04,nw-knnbr-04} for one-dimensional
data, while maintaining
the $O(\log n)$ memory size and expected query cost of 
traditional skip graphs~\cite{as-sg-03}
and SkipNet~\cite{hjstw-ssonp-03}.
We also introduce a bucketed version of our skip-web structure, which
improves the overall space bounds of our structure, while also
significantly improving the expected query and update times.
Indeed, if the memory size $M$ of our hosts is $O(n^{\epsilon})$, for
constant $\epsilon>0$, then our methods result in expected
constant-cost methods for performing queries and updates.
In Table~\ref{tbl-compare}, we highlight how our methods compare with
previous solutions for one-dimensional nearest-neighbor searches.

In addition to improving previous randomized distributed data
structures for one-dimensional data, 
we also extend the areas of applicability for such
structures, by introducing a general framework for constructing
distributed structures for multi-dimensional data sets, including
$d$-dimensional point sets, sets of character strings over fixed alphabets,
and (trapezoidal) maps of planar subdivisions defined by disjoint line
segments in the plane (e.g., as would be created by a campus or city map
in a geographic information system).
Our approach is based on viewing the nodes and links of 
an underlying data structures as having ranges (that is, sets) associated
with them.
By then applying a random selection process to these range spaces we
produce a distributed data structure (with low congestion) that is
able to perform efficient queries on the underlying structure.
For example, we can locate a point in a distributed two-dimensional
quadtree of $n$ points using only $O(\log n)$ messages, even if the
underlying quadtree has $O(n)$ depth,
providing a distributed analogue to the skip quadtree data structure
of Eppstein {\it et al.}~\cite{egs-sqsdd-05}.

\section{The Skip-Web Framework}
The skip-web framework applies to a wide variety of data querying
scenarios. In this section, we define the conditions that a data querying
scenario must satisfy in order to allow for a skip-web implementation.
Intuitively, our conditions apply to any data structure built
deterministically from an input set $S$ using nodes and links between
those nodes.
Our scheme builds a hierarchical, distributed structure 
in a way that allows for fast queries.

\subsection{Range-Determined Link Structures}
We consider the input set of $n$ items
to be a ground set $S$ of items taken from some universe $U$.
In order for the skip-web framework to apply, we require
that $S$ and $U$ define a unique \emph{link structure}~\cite{dsst-mdsp-89}.
That is, the universe $U$ and the elements of $S$ 
determine a unique data structure $D(S)$
consisting of \emph{nodes} and \emph{links} connecting pairs of those nodes.
Furthermore, we assume that
each node and link in $D(S)$ is associated with a range (that is, a set) 
of values from $U$ and
there is an incidence between a node and a link if and only if
their respective ranges have a non-empty intersection.

For example, if $U$ is a one-dimensional total order and $S$ is a 
finite subset of $U$, then 
$D$ could be an ordered linked list on the elements
from $S$.
In this case, the range associated with each node would 
be a unique element
from $S$ and the range associated with a 
link joining two nodes storing $x$ and $y$, respectively, would be
the interval $[x,y]$.
Such a linked list $D(S)$ is therefore a 
unique link structure with the incidences of
its nodes and links being defined by their associated ranges.
In general, we call such a structure
a \emph{range-determined} link structure.


As another example, consider $D(S)$ as
a digital trie for a set $S$ of $n$ character
strings, defined for a fixed alphabet.
The range associated with 
a node $v$ in $D(S)$ would naturally correspond to the singleton set
containing the string that leads to $v$ in $D(S)$.
Likewise, the range associated with
an edge $(v,w)$ in $D(S)$ would naturally correspond to the set of
all strings of the form $xy$, where $x$ is the string that leads to $v$ and 
$y$ is a (possibly empty) prefix of the string associated with the edge
$(v,w)$.
Thus, the intersection relationships between these sets of strings 
gives rise to the incidences between the nodes and links in $D(S)$.

\subsection{Defining a Set-Halving Lemma}
Let $S$ be a ground set of $n$ items taken from some universe 
$U$, and let $D(S)$ be 
the range-determined link structure for $S$.
Further, let $T$ be a subset of $S$ and let $D(T)$ be the range-determined
link structure for $S$ (defined by the same data-structure construction
function, $D$).

\begin{figure*}[tb]
\centerline{\includegraphics[scale=0.65]{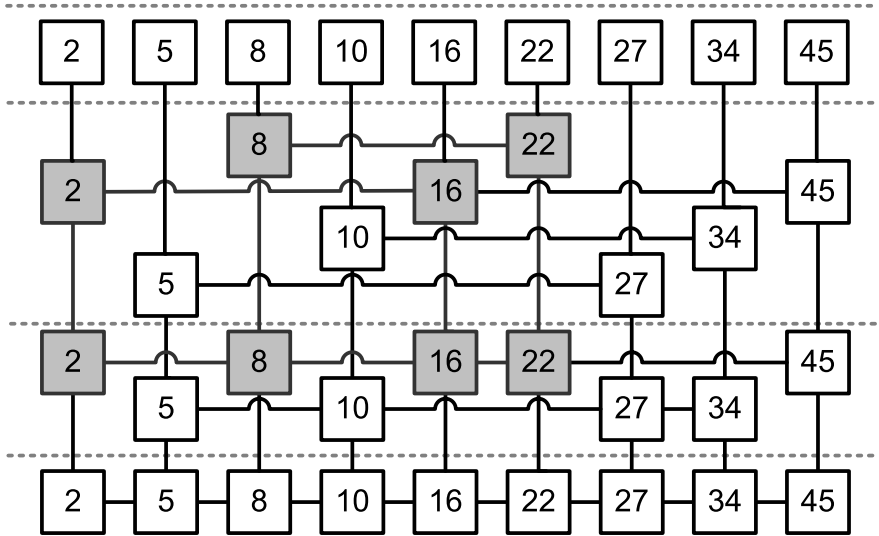}}
\vspace*{-10pt}
\caption{A one-dimensional skip-web. Each node is copied to the next
level $0$-list or $1$-list, respectively, with probability $1/2$.
The different levels of the skip-web are separated by dashed lines.
Note that if we follow pointers down from any top-level node, the
structure ``looks'' like a skip-list.
The nodes shown in gray could belong to a single host in the
skip-web.
(A skip-graph instead stores ``towers'' of nodes with the
same key at the same host.)}
\label{fig:blocking}
\end{figure*}

We say that a range $Q$ for a node or link in $D(T)$ 
\emph{conflicts} with a range $R$ for a node or link in
$D(S)$ if $Q\cap R\not=\emptyset$ (it is possible that $Q=R$, which we still
count as a conflict).
For a given range $Q$ for $D(T)$, we let $C(Q,S)$ denote the set of all
ranges for $D(S)$ that conflict with $Q$.
We call this set the \emph{conflict list} for $Q$.

The range-determined link structures we consider in this
paper all have an important property with
respect to the way that ranges conflict when we randomly halve
a given ground set.
In particular, 
we utilize the following:

\begin{itemize}
\item
\emph{Template for a Set Halving Lemma:}
Given a set $S$ of $n$ items taken from $U$,
let $T$ be a subset of $n/2$ items chosen uniformly at random from $S$.
If there exists a constant $c>0$ such that 
we can show that $E(|C(Q,S)|)\le c$, for any item $q\in U$ and for the maximal range $Q$ of $D(T)$ containing $q$, then we say
that $U$ and $D$ have a \emph{set halving lemma}.
\end{itemize}

We demonstrate set halving lemmas for a number of range-determined link
structures, including:
\begin{enumerate}
\item compressed quadtrees and octrees for points in $\R^d$
\item compressed digital tries for character strings over fixed alphabets
\item linked lists of sorted sets
\item Trapezoidal diagrams of non-crossing sets of line segments in the plane.
\end{enumerate}
For example, we have the following:

\begin{lemma}
\label{lemma:1dset}
Let $U$ be a total order, let $S$ be a set of $n$ items from $U$, let $q$ be an arbitrary item in $U$, and
let $T$ be a subset of $n/2$ items chosen uniformly at random from $S$.
Further, let $D$ be the range-determined linked structured defined by a
doubly-linked list, and let $Q$ be the maximal range in $D(T)$ containing $q$.
Then $E(|\C(Q,S)|)$ is $O(1)$.
\end{lemma} 
\begin{proof}
For any $x$ in $S$, the probability that $x$ belongs to $Q$ is $2^{-|[q,x)\cap S|}$.  Thus, $E(|Q\cap S|)$ is a sum of inverse powers of two, in which each power appears at most twice; therefore, $E(|Q\cap S|)\le 4$. $|C(Q,S)|\le 2|Q\cap S|-1$,
so $E(|C(Q,S)|)\le 2E(|Q\cap S|)-1<=7$.
\end{proof}

This is the set halving lemma for a sorted linked list.
Rather than immediately showing
that the remaining range-determined link structures
above also have
set halving lemmas, let us continue our discussion with the way 
a range-determined link structure with a set
halving lemma can be used to construct an efficient distributed data
structure.
We use the sorted linked list as a running example, but the
construction applies to other link structures as well.

\subsection{Skip-Web Levels}
Suppose, then, 
that we are given a set $S$ of $n$ elements, taken from a 
universe $U$, with a range-determined link structure $D$ with 
a set-halving lemma.
We define $S_0=S$ to be the level-0 subset of $S$ and we let $D(S_0)$ be the
level-0 structure for $S_0$.
For example, with a set $S_0$ taken from a total order, $D(S_0)$ is the
sorted linked list defined on $S_0$.

Suppose we have inductively defined a level-$i$ 
set $S_b$, where $b$ is an $(i+1)$-bit binary string.
We generate a new random bit for each $x$ in $S_b$.
If the bit for $x$ is $0$, then we include $x$ in a set, $S_{b0}$;
otherwise, we include $x$ in a set, $S_{b1}$.
For each range $Q$ for $D(S_{b0})$ we store hyperlink pointers to the 
nodes and links for 
the ranges in $\C(Q,S_b)$, that is, the ranges 
in $D(S_b)$ that conflict with $Q$.
In this case,
a pointer consists of a pair $(h,a)$, where $h$ is the ID of a host and $a$
is an address on that host where the item being referred to is stored
(we assume that links have an address).
Likewise,
for each range $Q$ in $D(S_{b1})$ we store hyperlink pointers to the 
ranges in $\C(Q,S_b)$, i.e., in $D(S_b)$, that conflict with $Q$.
We repeat this process until we define sets 
with $\lceil\log n\rceil$-bit indices.
The expected size of each top-level structure is $ O(1) $, 
if the total number of levels is $O(\log n)$.
(See Figure~\ref{fig:blocking}.)

\subsection{Distributed Blocking}
Given a range-determined link structure and a ground set $S$ with a
set-halving lemma, the final ingredient for constructing a skip web is to
assign the nodes and links of the various levels to hosts in the network.
For now, let us simply assume that this assignment is arbitrary; so that each
host on the network gets $O(M)$ nodes and links from among the $O(n\log n)$
possible.
So, for example, if $M=\log n$, then we can use $n$ hosts to store the data
structure using $O(\log n)$ space per node.
This allocation, and the generality of the skip-web approach,
leads to fast query times for several multi-dimensional distributed search
problems, which we detail in Section~\ref{sec:applications}.

\subsubsection{Improved Blocking for One-Dimensional Data}
For one-dimensional data, the above arbitrary blocking 
approach would match the performance of
skip-graph~\cite{as-sg-03} and SkipNet~\cite{hjstw-ssonp-03} structures.
We can improve this blocking strategy for one-dimensional data, however,
by a more clever approach, thereby 
constructing a distributed data structure
for one-dimensional data that is more efficient than skip-graphs and SkipNet
and faster than family trees~\cite{zh-ftaod-04}.
As with these other structures,
the queries our structure supports are 
one-dimensional nearest-neighbor queries,
which is equivalent to a one-dimensional point-location query,
that is, given a point $x$ in $U$, find a range $R$ for a node or link
in $D(S)$ that contains $x$.
For any universe $U$ of one-dimensional data
and link structure $D$ based on a doubly-linked list,
we desire that the
expected number of messages to answer a range-containment query is 
$O(\log n/\log M)$, where  $M$ is the memory size of each host.

Suppose, then, 
that we are given a set $S$ of $n$ elements taken from a 
one-dimensional universe $U$, and we use an ordered doubly-linked list $D$ 
as our range-determined link structure.
By Lemma~\ref{lemma:1dset}, this set and structure have a set-halving lemma.
As in the general case,
we define $S_0=S$ to be the level-0 subset of $S$ and we let $D(S_0)$ be the
level-0 structure for $S_0$, which, in this case, is a simple linked list
on $S_0$.

Recall that we inductively defined a level-$i$ 
sets $S_b$, where $b$ is an $(i+1)$-bit binary string using random sampling.
We store the data structure nodes and links of each $D(S_b)$ in a
hierarchical, stratified fashion.
We define level $i$ set as \emph{basic} if $i$ is a multiple of
$L=\lceil\log M\rceil$.
If $S_b$ is a basic set, then we divide $D(S_b)$ into $M/L$ blocks of
\emph{contiguous} ranges, that is, a contiguous portion of the nodes and links
in the linked list $D(S_b)$.

Let $\cal R$ be the contiguous set of ranges
of a basic $D(S_b)$ stored on some host $h$.
We also store at $h$ all the ranges of $D(S_{b0})$ that conflict with ranges
in $\cal R$ as well as the ranges of $\P(S_{b1})$ that conflict with ranges
in $\cal R$.
In fact, we store at $h$ the ranges in $D(S_{b00})$, $D(S_{b01})$,
$DP(S_{b10})$, and $D(S_{b11})$ that conflict with ranges on the next lower
level, and we continue this process for all the non-basic levels above $i$.
(Note that copies of some of these ranges may be stored on multiple hosts,
but these overlapping copies will only increase the total space by a
constant factor, since these sets of ranges are contiguous and
compact.)
This construction requires space proportional to $M$, as required.
The number of hosts needed is $H\le cn\log n/M$, for some constant $c$.
Thus, $H$ is $n$ if we choose $M$ to be $O(\log n)$.
The in-degree and out-degree of each nodes is $O(M)$, as well.

\subsection{Answering Queries}
To answer a query $q$ (which may be over multiple attributes) 
in a skip-web we begin with the ``root'' nodes and links
of a level-$k$ $D(S_b)$ stored on
the host originating the query.
We search in $D(S_b)$ as far as we can for $q$, stopping at a node or link
whose range $R$ includes $q$ or intersects $q$.
We then follow the hyperlinks for $R$ to the nodes of the level-$(k-1)$
structure, possibly stored at another host.
If the hyperlinks are stored on our current host, then we follow these
connections internally.
Otherwise, following these hyperlinks is done by sending a message for $q$ to
the host(s) storing the hyperlinks, asking the host(s) to process the query
as far as they can internally and then return either an answer or hyperlinks
to other hosts at which we can continue the search.
We continue following all the active hyperlinks in this way, possibly sending
messages to other hosts, until we complete all the processing
for the query $q$.

The key observation for analyzing the message complexity for answering such a
query is to note that
we have to consider an expected constant number of conflicting ranges in
going from one level to the next (by following 
hyperlinks leading from this level to
ultimately yield answers) will be processed,
e.g., just one hyperlink in a one-dimensional nearest-neighbor search.
In the case of general skip-web structures, this implies a query time of
$O(\log n)$ message per answer.
In the case of bucketed one-dimensional data, we note that we need follow an
expected constant number of external hyperlinks
(to new hosts to determine which pointer to follow in our search) 
for basic levels.
The pointers we need to follow for non-basic levels need not branch out to
other hosts, by construction.
Thus, for example, if $M$ is $O(\log n)$, then we can answer
nearest-neighbor queries using $O(\log n/\log\log n)$ 
expected message complexity.
We summarize:

\begin{theorem}
Given a set $S$ of $n$ elements taken from a 
universe $U$,
and having a range-determined link structure with a set-halving lemma,
then we can construct a distributed
skip-web structure that uses $n$ hosts, with memory
size $O(\log n)$, congestion $O(\log n)$, 
and query complexity $O(\log n)$, which can be improved to
$O(\log n/\log\log n)$ for one-dimensional data.
\end{theorem}

For general values of the host memory size, $M$, 
when we are
storing one-dimensional data, we get a structure we call the 
\emph{bucket skip-web}, whose performance is as mentioned
in Table~\ref{tbl-compare}.

\begin{figure*}[tb]
\centerline{\includegraphics[scale=0.8]{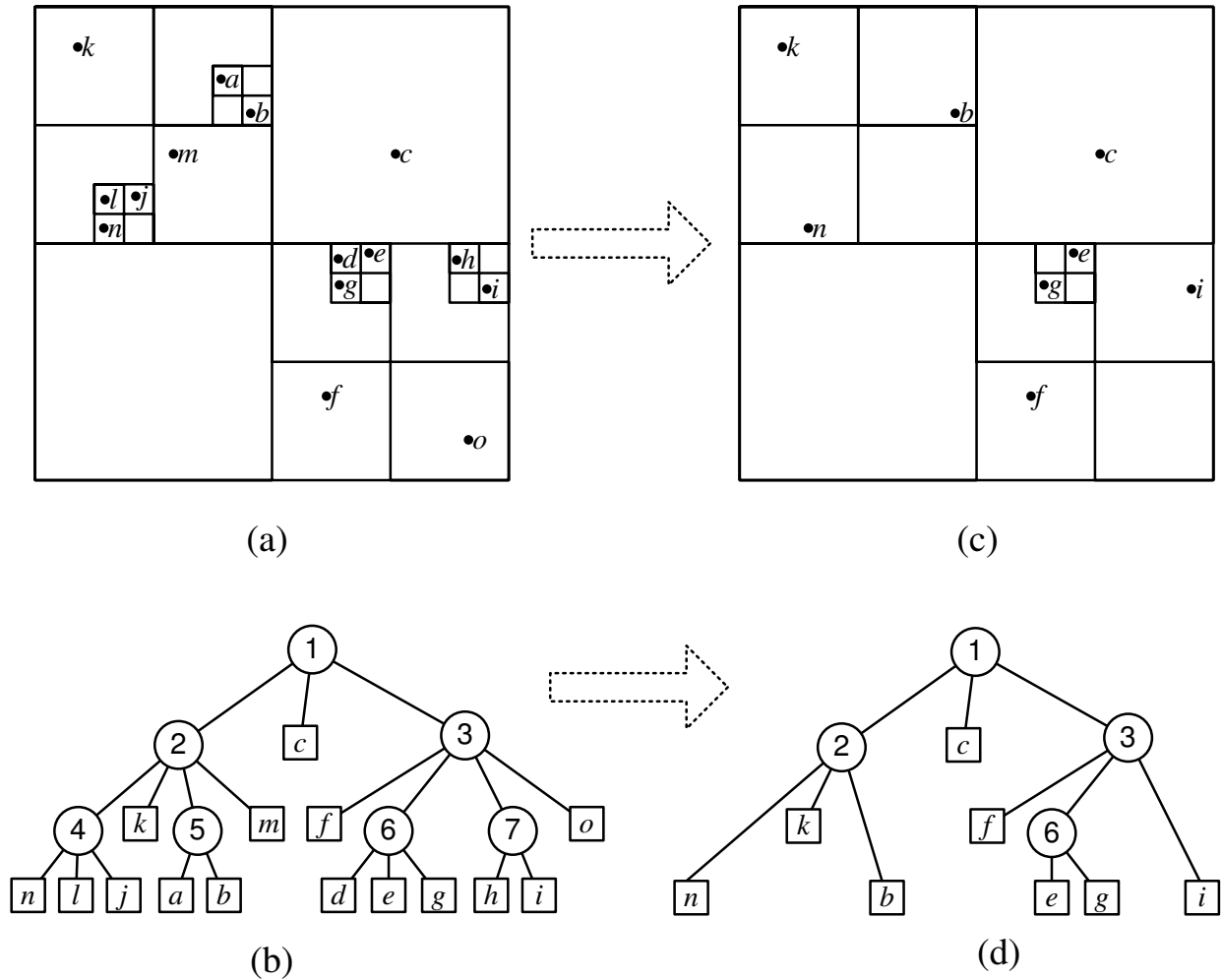}}
\caption{Illustrating the halving lemma for quadtrees.
(a) the regions for a compressed quadtree defined on a set $S$ of points;
(b) the compressed quadtree for the points in $S$;
(c) an example random subset $T$ of the points in $S$ and their
associated quadtree regions;
(d) the compressed quadtree for the points in $T$.}
\label{fig:skipquad}
\end{figure*}

\section{Multi-Dimensional Skip Webs}
\label{sec:applications}
Having discussed the general skip-web framework and how to 
optimize it for one-dimensional data, 
we discuss in this section how to construct distributed
skip webs for specific multi-dimensional data sets.

\subsection{$d$-Dimensional Point Sets: Quadtrees and Octrees}
In this section, we outline a construction of a skip-web for 
compressed quadtrees and octrees in $d$-dimensional space, for any fixed
constant $d\ge2$.
A quadtree (for two-dimensional data) or octree (for higher dimensions) is
defined in terms of a set $S$ of $n$ points and a bounding hypercube 
(a square in two-dimensions).
The root $r$ of the tree is associated with the bounding cube.
This cube is subdivided into $2^d$ subcubes with side-length half of that of
the bounding cube and each non-empty subcube is a child of $r$ and is
the root of a tree that is recursively subdivided in the same manner.
This tree is then converted into a compressed version by compressing chains
of nodes with only one child into edges.
This tree has $O(n)$ nodes and links, but can have depth $O(n)$
in the worst case.
More importantly, for the context of this paper, a compressed quadtree or
octree is a range-determined link structure.

By showing that quadtrees and octrees have a set-halving lemma, 
we show that we can build a skip-web structure for multi-dimensional point
sets that can perform point-location in the subdivision of space 
defined by the hypercubes associated with nodes in the tree.
This point location can be done with $O(\log n)$ messages and, as 
Eppstein {\it et al.}~\cite{egs-sqsdd-05}
show, such point location
queries can be used to answer approximate nearest-neighbor queries
and approximate range searches in $d$-dimensional space.
In the case of defining a range-determined link structure from a 
quadtree or octree, the range associated with each node is
its associated cube and the range associated with a link is that of the child
node for that link.
We can define a set-halving lemma for quadtrees and octrees as
follows.
(See Figure~\ref{fig:skipquad}.)

\begin{lemma}
Let $S$ be any set of $n$ points in $d$-dimensional space, 
for fixed constant $d\ge 2$,
let $T$ be formed by choosing
each member of $S$ independently with probability $1/2$, and let $C$ be a
(hyper)cube associated with a node in the quadtree/octree $D(T)$.
Then the expected number of node hypercubes
in $D(S)$ that conflict with $C$ is $O(1)$.
\end{lemma}

\begin{proof}
Follows from results from Eppstein {\it et al.}~\cite{egs-sqsdd-05}.
\end{proof}

This lemma and the skip-web framework
implies that we can perform point-location queries in 
a network of $n$ hosts with $O(\log n)$ messages (even if the underlying
quadtree or octree has depth $O(n)$).

\subsection{Character Strings: Tries}
We have already discussed above 
that a compressed trie defined by a set $S$ of 
$n$ character strings defined over
a fixed alphabet is a range-determined link structure.
In this section, we show that tries defined over such a set $S$ has a
set-halving lemma.

\begin{figure*}[tb]
\centerline{\includegraphics[scale=0.85]{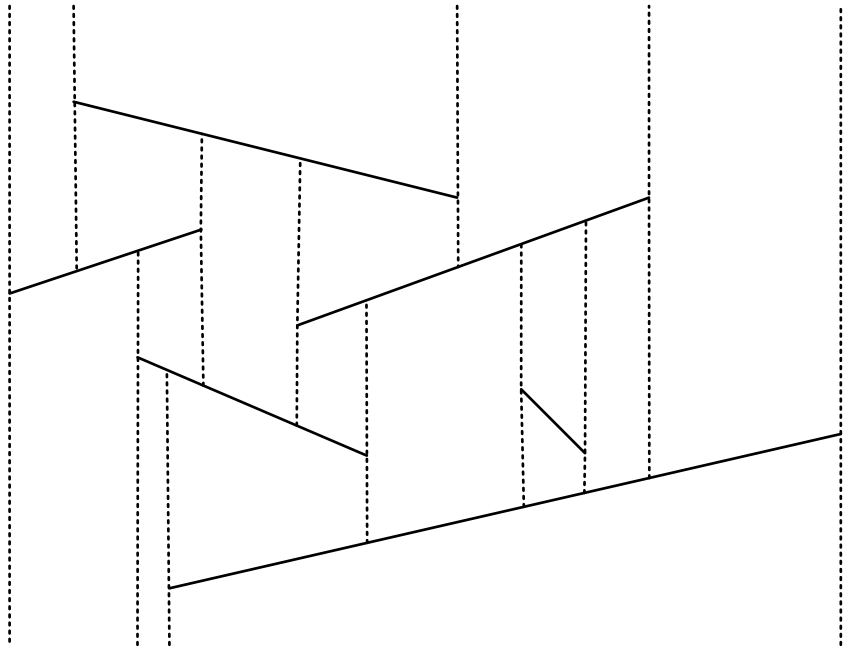}}
\caption{An example trapezoidal map.}
\label{fig:trapezoids}
\end{figure*}

\begin{lemma}
Let $S$ be any set of $n$ character strings over a fixed alphabet,
let $T$ be formed by choosing
each member of $S$ independently with probability $1/2$, and let $s$ be a
set of strings associated with a node or link in $D(T)$.
Then the expected number of ranges (i.e., nodes and links)
in $D(S)$ that conflict with $s$ is $O(1)$.
\end{lemma}

\begin{proof}
Consider an edge of $D(T)$.  Each such edge corresponds to a
path $P$ in $D(S)$, which could just be the same edge.
Note that if $P$ is more than a single edge, then each node along $P$
branches to at least one string in $S-T$.  Moreover, the string(s)
for any such node are disjoint from the string(s) for any other node missing 
from $D(T)$ but in $D(S)$.
Since each string has a probability of $1/2$ of being in $S-T$, this
implies that the expected number of nodes along $P$ is $O(1)$.
\end{proof}

Applying this lemma to the skip-web framework implies that we can perform
trie searches for an arbitrary character string using $O(\log n)$ messages,
even if the underlying trie has depth $O(n)$.
The subqueries performed by each host in this case involves finding the first
place where a query substring differs with the string defined by a substring
associated with a link in a trie.

\subsection{Trapezoidal Maps}
In this section, we outline a construction of a skip-web for
trapezoidal maps.  
A \emph{trapezoidal map} $D(S)$ 
for a set $S$ of non-crossing line segments in the plane
is a subdivision of the plane defined by the input segments and additional
segments formed by extending vertical rays up and down from each segment
endpoint until it hits another segment or extends to infinity.
(See Figure~\ref{fig:trapezoids}.)
As with our other skip-web applications,
we begin with a
set-splitting lemma. 

\begin{lemma}
Let $S$ be any set of disjoint line segments, let
$x$ be any point in the complement of $S$, let $T$ be formed by choosing
each member of $S$ independently with probability $1/2$, and let $t$ be the
trapezoid containing $x$ in $D(T)$.  Then the expected number of trapezoids
in $D(S)$ that conflict with $t$ is $O(1)$.
\end{lemma}

\begin{proof}
First note that, for any trapezoid $t$ in $D(T)$, 
the number of conflicts is exactly
$1 + a + 2b + 3c$, where
$a$ is the number of segments of $S$ that cut all the way across $t$
and have no endpoints interior to $t$,
$b$ is the number of segments of $S$ with one endpoint interior to $t$,
and
$c$ is the number of segments of $S$ with both endpoints interior to $t$.
This equation can be shown by induction on $|S-T|$: if $|S-T|=0$, 
the number of conflicts is $1$ ($t$ conflicts with itself) and 
each additional segment added to get from $T$ to $S$ 
increases the number of
conflicting trapezoids by one (if it cuts $t$ and has no vertex in $t$),
by two (if it has one vertex in $t$), or
three (if it has both vertices in $t$).

To show that the expected values of $a$, $b$, and $c$ are $O(1)$, let us
consider bounding $a$ (the methods for $b$ and $c$ are similar).
Given a trapezoid $t$ of $D(T)$,
note that in order for $t$ to exist in $D(T)$ none of the segments of $S$
that intersect the interior of $t$ could have been chosen to be in $T$.
The expected value of $a$ for $t$, therefore, can be bounded by
{\small
\[
\sum_{i=0}^{n} i 
	\Pr(\mbox{none of $i$ segments cutting $t$ were chosen to be in $T$}).
\]
}
Since each segment of $S$ has probability $1/2$ of being chosen to be in $T$,
this bound is at most
\[
\sum_{i=0}^{n} \frac{i}{2^i},
\]
which is $O(1)$.
\end{proof}

\pagebreak
\section{Updates in a Skip-Web}
Having described how a skip-web structure can be used for efficiently performing
query routing, we describe in this section how to efficiently perform
updates in a skip-web.  
For the sake of simplicity, we restrict our attention
here to the insertion of a new element of the ground set $S$,
assuming that there is only a single update being performed at a
time, that all updates eventually complete, and that queries are
temporarily blocked at nodes being updated (until the update
completes).
Deletion is similar (and actually simpler).

Suppose some host $h$ in our peer-to-peer network wishes
to insert a new element $x$ of the ground set $S$.
We begin by performing a search to locate
$x$ in the level-$0$ structure, $D(S)$.  
We then update this structure to construct $D(S\cup \{x\})$.
For one-dimensional linked lists, quadtrees, octrees, and tries, this
update involves the insertion of $O(1)$ nodes and links.
For a trapezoidal map, it could involve 
the creation of an output-sensitive number of 
new nodes and links, however, so let us assume for this application
that we are considering the insertion of a new trapezoid (which we
would amortize against the output-sensitive term in an insertion
bound).
So let us assume that the number of new nodes and links created by
the insertion is $O(1)$.

Having added the new nodes and links to go from 
$D(S)$ to $D(S\cup \{x\})$, we then generate $\lceil\log n\rceil$
random bits, which will determine which higher-level structures $x$
should be added to.
We add $x$ to the structures for these levels in a bottom-up fashion,
starting from the nodes and links conflicting with the $O(1)$ nodes
and links that we replaced when adding $x$ to $S$.
Given these conflicts, we can ``move up'' to the higher-level
structure, add new nodes and links for the insertion of $x$, and then
continue the bottom-up procedure.
By our assumption, the number of messages needed to update each level
is $O(1)$; hence, the expected message complexity is $O(\log n)$.
In fact, in the case of one-dimensional data, the expected message complexity
is $O(\log n/\log\log n)$, since we only need to send new messages to
the structures on basic levels (any block splits we need to make as
the result of an insertion can be amortized against $O(\log n)$
previous insertions that led to that split).


\subsection*{Acknowledgments} This research is supported in part by
the NSF under grants CCR-0098068, CCR-0225642, and CCR-0312760.

{\small
\bibliographystyle{abbrv}
\bibliography{geom,crypto,extra,goodrich}
}

\end{document}